\documentclass{arxiv}

\graphicspath{{figs/}{figures/}{pictures/}{images/}{./}} 
\usepackage[colorlinks,urlcolor=blue,linkcolor=blue,citecolor=blue]{hyperref}
\usepackage{booktabs}
\usepackage{balance}
\usepackage{fontawesome5}
\usepackage{xcolor}

\usepackage{mdframed}

\newcommand{\papertitle}{Automating the Path: An R\&D Agenda for Human-Centered AI and Visualization}

\jname{Author Preprint}
\jtitle{\papertitle}
\pubyear{2025}

\newcommand{\rev}[1]{{#1}} 

\definecolor{BlueGray}{HTML}{546E7A}
\definecolor{BlueGrayLight}{HTML}{607D8B}
\definecolor{Yellow}{HTML}{FBC02D}
\definecolor{Amber}{HTML}{FF8F00}
\definecolor{Orange}{HTML}{FB8C00}
\definecolor{DeepOrange}{HTML}{FF7043}
\definecolor{Indigo}{HTML}{3949AB}
\definecolor{Teal}{HTML}{00897B}
\definecolor{LightGreen}{HTML}{7CB342}
\definecolor{LightBlue}{HTML}{0277BD}
\definecolor{Purple}{HTML}{AB47BC}
\definecolor{DeepPurple}{HTML}{7E57C2}

\begin{document}


\sptitle{AUTHOR PREPRINT}

\title{\papertitle}

\author{Niklas Elmqvist}
\affil{Aarhus University, Aarhus, Denmark}

\author{Clemens Nylandsted Klokmose}
\affil{Aarhus University, Aarhus, Denmark}

\markboth{GRAPHICALLY SPEAKING}{GRAPHICALLY SPEAKING}
\markboth{AUTHOR PREPRINT}{AUTHOR PREPRINT}

\begin{abstract}
    The emergence of generative AI, large language models (LLMs), and foundation models is fundamentally reshaping computer science, and visualization and visual analytics are no exception.
    We present a systematic framework for understanding how human-centered AI (HCAI) can transform the visualization discipline.
    Our framework maps four key HCAI tool capabilities---amplify, augment, empower, and enhance---onto the four phases of visual sensemaking: view, explore, schematize, and report.
    For each combination, we review existing tools, envision future possibilities, identify challenges and pitfalls, and examine ethical considerations.
    This design space can serve as an R\&D agenda for both visualization researchers and practitioners to integrate AI into their work as well as understanding how visualization can support HCAI research.
\end{abstract}

\maketitle


\chapteri{A}lmost \rev{exactly two decades ago, the late Jim Thomas together with Kristin Cook edited \textit{Illuminating the Path: A Research and Development Agenda for Visual Analytics}~\cite{thomas2005illuminating}, a book that helped give rise to the research field of visual analytics (VA).
Written by a multidisciplinary panel of some 30 researchers and practitioners in visualization, databases, analytics, psychology, and machine learning during an intense year of workshops, training, and ideation, the book---while clearly representing a very U.S.-centric vision---has nevertheless served to establish visual computing and analytics in the greater international scientific community.}

\rev{Twenty years later, visual analytics is a familiar and accepted part of the visualization field, almost to the point that the two terms are interchangeable. 
However, while what today might broadly be called \textit{artificial intelligence}---computational science, unsupervised algorithms, and machine learning---have been part of visual analytics from the start, progress in AI since then has been nothing short of revolutionary. 
Deep learning, large language models (LLMs), and generative AI are fundamentally reshaping computer science, and the visualization and visual analytics area is no exception.
But does this mean that yet another new discipline is needed, one that builds on and supersedes visual analytics by adopting AI to an even greater extent?
Put differently, do we need a new research agenda that can reilluminate the path twenty years after the original?}

In this paper, we argue that no such new discipline is needed but that the existing field of visualization and visual analytics is sufficient to harness these revolutionary AI advances while maintaining its core human-centered principles.
Rather than an entirely new discipline, we argue that \rev{only a course correction is needed;} that instead of merely illuminating the path, deeper integration of AI into our fundamentally human-centered approach to sensemaking will enable us to ``automate'' our progress along this path.
This requires adopting the emerging discipline of \textit{human-centered AI} (HCAI)~\cite{Shneiderman2022}---the combination of artificial intelligence capabilities with human-computer interaction principles to create AI systems that augment human intelligence, respect human values, and safeguard user agency---where visualization and visual analytics stand poised to play a decisive role~\cite{DBLP:Hoque2024}.

Towards this end, we present a design space that maps \rev{four key HCAI capabilities}---amplify, augment, empower, and enhance---onto the four phases of visual sensemaking: prepare, explore, schematize, and report.
Each combination of HCAI capability and sensemaking phase gives rise to interesting examples for how AI can support visualization and vice versa: current tools, future potential, challenges and pitfalls, as well as ethical considerations. 
This approach also mirrors Thomas and Cook's original vision~\cite{thomas2005illuminating} for supporting analytical reasoning, harnessing visual representations and interactions, transforming and processing data, and supporting its production, presentation, and dissemination.
We see this design space as a new research and development agenda for how visualization and visual analytics can approach the next 20 years of progress following an illuminated path.

\begin{mdframed}[backgroundcolor=blue!10,
    frametitle={\section*{\textcolor{white}{Sidebar: Illuminating the Path}}},
    frametitlerule=true, frametitlebackgroundcolor=bgcolor]

\textit{Illuminating the Path: An R\&D Agenda for Visual Analytics} (2005) by James J.\ Thomas and Kristin A.\ Cook~\cite{thomas2005illuminating} was a seminal work that established visual analytics as a distinct scientific discipline.
Published in the wake of the 9/11 terrorist attack in the United States, it outlined the first research agenda for visual analytics as part of the U.S.\ Department of Homeland Security National Visualization and Analytics Center (NVAC).

The book established visual analytics as a multidisciplinary field, provided the first formal definition of visual analytics as ``the science of analytical reasoning facilitated by interactive visual interfaces,'' and influenced a generation of visualization research and homeland security applications.
Its core principles remain relevant to modern visual analytics systems.
The book introduced what became the motto of visual analytics: ``\textit{Detect the expected and discover the unexpected.}''
Here we list its main chapters:

\begin{enumerate}
    \item[\textbf{Ch.1}] Grand Challenges
    \item[\textbf{Ch.2}] The Science of Analytical Reasoning
    \item[\textbf{Ch.3}] Visual Representations \& Interactions
    \item[\textbf{Ch.4}] Data Representations and Transformations
    \item[\textbf{Ch.5}] Production, Presentation \& Dissemination
    \item[\textbf{Ch.6}] Moving Research Into Practice
    \item[\textbf{Ch.7}] Positioning for Enduring Success 
\end{enumerate}
\end{mdframed}

\section[Background]{BACKGROUND}

In many ways, visualization and AI represent polar opposites of the relationship between humans and computers~\cite{DBLP:journals/pnas/Heer19}: the former is fundamentally part of the \textit{intelligence augmentation} (IA) movement, which sees computers as tools designed to enhance human capabilities, whereas AI, on the other hand, strives to create autonomous entities simulating human intelligence. 
Despite parallel and mostly independent evolution, visualization and visual analytics has long maintained ties with the AI side; Endert et al.~\cite{DBLP:journals/cgf/EndertRTWNBR17} reviewed how classic machine learning techniques such as dimension reduction, clustering, classification, and regression are core components of many visual analytics workflows in areas such as text analytics, multimedia, streaming data, and bioinformatics.
However, they also point out that many steerable ML algorithms exist that would be advantageous to embed into visual analytics, but with the caveat that this could lead to automation surprise, decreased trust, and poor interpretability.

Wang et al.~\cite{DBLP:journals/tvcg/WangCWQ22} built on this idea by surveying the use of machine learning for what they call visualization-related problems: data processing, visual mapping, insight communication, style imitation, guided interaction, user profiling, and interpreting visualizations.
We adopt their pipeline-centric view in this paper, but reduce their seven processes down to a set of four.

The rise of \textit{deep learning}---artificial neural networks with multiple layers that can automatically learn representations and patterns---and \textit{generative AI}---automatic systems that can create new content, such as text, images, code, or music, by learning patterns from existing data---is threatening to upend this relationship all over again.
Ye et al.~\cite{DBLP:journals/vi/YeHHWXLZ24} surveyed the state of the art from 81 papers based on four stages---data enhancement, visual mapping generation, stylization, and interaction---for the purpose of identifying challenges and future research opportunities.
Finally, Basole and Major~\cite{DBLP:journals/cga/BasoleM24} studied potential roles for generative AI in seven stages of a visualization workflow through the lens of three complementary capabilities: creativity, assistance, and automation. 
Compared to Ye et al., Basole and Major's focus is less on a literature survey and more a call to action on research for concrete visualization tasks where these capabilities can help.

Our approach in this paper draws on the above prior art and has a similar four-stage pipeline, but introduces \rev{the lens of human-centered AI (HCAI)}~\cite{Shneiderman2022} for visualization proposed by Hoque et al.~\cite{DBLP:Hoque2024} to speculate on AI amplification, augmentation, empowerment, and enhancement for the entire data analytics workflow from ETL to dissemination.

\section[Combining Visualization and AI]{COMBINING VISUALIZATION \& AI}

\rev{\textit{Visualization-enabled HCAI tools}~\cite{DBLP:Hoque2024} use interactive visualization to facilitate a human-centered approach to AI.}
This definition builds in turn on Shneiderman's four \textit{capabilities} of ``AI-infused supertools'' from his book on \textit{Human-Centered AI}~\cite{Shneiderman2022}.
We enumerate them below along with our interpretation of their meaning~\cite{DBLP:Hoque2024}. 

\begin{itemize}
    \item[\faVolumeUp] \rev{\textbf{Amplify}: strengthening existing abilities.}

    \item[\faPlusCircle] \rev{\textbf{Augment}: adding new abilities.}
    
    \item[\faFistRaised] \rev{\textbf{Empower}: enabling tasks previously impossible.}
    
    \item[\faBolt] \rev{\textbf{Enhance}: improving existing abilities.}
    
\end{itemize}

We can envision many ways in how these capabilities can be realized depending on the subject domain. 
For visualization, it makes sense to think of roughly four separate activities based on the \textit{sensemaking phases}~\cite{DBLP:conf/chi/RussellSPC93} where HCAI tools can be beneficial: 

\begin{itemize}
    \item[\faTools]\textbf{Prepare}: transform raw data into structured, standardized formats through preprocessing, wrangling, and quality assurance.

    \item[\faCompass]\textbf{Explore}: investigate patterns, relationships, and anomalies in the data through interactive analysis and visualization.
    
    \item[\faSitemap]\textbf{Schematize}: organize insights into coherent structures through progressive sensemaking and well-defined hierarchies of information.

    \item[\faFile*]\textbf{Report}: make conclusions, communicate findings, and support decision-making through visual narratives that address stakeholder needs.

\end{itemize}

In Table~\ref{tab:design-space}, we describe HCAI tool capabilities for each of these phases in turn.
In the following sections, we give examples current HCAI tools as well as challenges, pitfalls, and ethical considerations for each of the phases.
We close the paper by proposing a joint R\&D agenda for visualization-enabled HCAI tools.

\begin{table*}[htb]
    \centering
    \caption{\textbf{Design space for visualization-enabled HCAI tools.}
    The four HCAI tool capabilities~\cite{DBLP:Hoque2024,Shneiderman2022} form a design space when combined with the four sensemaking phases~\cite{DBLP:conf/chi/RussellSPC93} typically defined for data analytics and visualization.
    In each cell, we give an example of how HCAI may support visualization for the respective combination.}
    \label{tab:design-space}
    \begin{tabular}{p{0.10\textwidth}>{\raggedright}p{0.2\textwidth}>{\raggedright}p{0.175\textwidth}>{\raggedright}p{0.175\textwidth}>{\raggedright}p{0.175\textwidth}}
        \toprule
        & \multicolumn{1}{c}{\textbf{Prepare}} & \multicolumn{1}{c}{\textbf{Explore}} & \multicolumn{1}{c}{\textbf{Schematize}} & \multicolumn{1}{c}{\textbf{Report}}
        \tabularnewline
        \midrule
        \textbf{Amplify}
        & Normalizing multiple and heterogeneous datasets
        & Support for hypothesis generation
        & Pattern detection and annotation support
        & Automated narrative construction
        \tabularnewline
        \midrule
        \textbf{Augment}
        & Generating artificial data
        & \rev{Chart recommendation}
        & \rev{Automated insight capture across tools}
        & Audience-specific storytelling support
        \tabularnewline
        \midrule
        \textbf{Empower}
        & \rev{Automated data wrangling}
        & \rev{Exploration feedback and bias detection}
        & Intelligent insight shoebox~\cite{DBLP:conf/chi/RussellSPC93}
        & Variable-length and -format reports
        \tabularnewline
        \midrule
        \textbf{Enhance}
        & \rev{Automated quality assurance and curation}
        & Analysis coverage management
        & Automated insight management
        & \rev{Detecting overclaims and misleading charts}
        \tabularnewline
        \bottomrule
    \end{tabular}
\end{table*}

\section[Prepare]{\faTools~PREPARE}

Data preparation encompasses the essential processes of extracting data from various sources, transforming it into consistent formats, and loading it into analysis systems (ETL).
This phase includes activities such as cleaning, standardizing, integrating multiple data sources, and ensuring data quality.
The \rev{challenges in this phase include} dealing with data quality issues, managing large-scale data transformations, and establishing reliable data pipelines while maintaining data provenance and reproducibility.

\begin{figure}[htb]
    \centering
    \includegraphics[width=\linewidth]{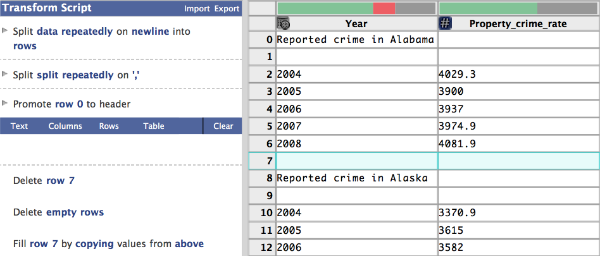}
    \caption{\textbf{Stanford Wrangler.}
        The left panel presents a transformation history tracking data provenance, a transform selection menu for manual ETL operations, and contextually suggested transformations based on current data selection.
        Interactive parameters within transform descriptions (shown in bold) enable direct manipulation of the ETL pipeline.
        The right panel displays the resulting data table with integrated data quality meters above each column, providing immediate feedback on transformation outcomes.
    }
    \label{fig:wrangler}
\end{figure}

\smallskip\textbf{Current HCAI Tools.}
The Stanford Wrangler~\cite{DBLP:conf/chi/KandelPHH11} (Figure~\ref{fig:wrangler}), which became the foundation for the commercial Trifacta Wrangler (acquired in 2022 by Alteryx), demonstrates how HCAI can support the prepare phase of visual sensemaking by combining automated transformation suggestions with interactive visualization.
The system uses programming-by-demonstration and inference techniques to automatically suggest relevant data transformations, while providing users direct manipulation controls and visual previews of transformation effects.
This approach helps analysts efficiently transform raw data into structured formats, validate data quality, and create reusable transformation scripts.
By automating complex transformations while maintaining human oversight, Wrangler shows how HCAI tools can significantly reduce the time and effort required for data preparation while ensuring transformation accuracy and reusability.

\smallskip\textbf{Challenges, Pitfalls \& Ethical Considerations.}
Decisions made during the data preparation phase---such as data cleaning, transformation, and integration---can introduce biases, obscure key patterns, or compromise the integrity of the resulting visualizations.

\begin{itemize}

\item \textit{Data loss:} Depending on an automated method to prepare the data may lead to some data being ignored; for example, unmatched rows in a join (e.g., customers without transactions) may be dropped, resulting in incomplete datasets.

\item \textit{Conversion errors:} Automatic data management may lead to conversion errors.

\item \textit{Semantic mismatch:} An overzealous automatic data wrangling method may end up structurally joining two tables that are not semantically compatible; for example, combining two census tables based on non-unique fields such as names.

\item \textit{Provenance and transparency:} Complex data transformations may cause the data provenance to be unclear; for example, a synthesized dataset combining many separate tables and search query results can be difficult to assess.

\item \textit{Bias amplification:} Automated data preparation methods may inadvertently amplify existing biases in the data; for example, automatic outlier removal might disproportionately affect minority groups in demographic data.

\item \textit{Privacy leakage:} Automation may inadvertently reveal personally identifiable information through correlation; for example, combining anonymized datasets that share common attributes.

\item \textit{Resource optimization:} Excessive reliance on AI for data preparation may lead to computational inefficiencies, particularly with large-scale datasets requiring real-time processing.

\end{itemize}

\section[Explore]{\faCompass~EXPLORE}

Exploratory data analysis represents the iterative process of visual and analytical discovery, where analysts generate and refine hypotheses through interactive visualization and analysis. 
This phase emphasizes the science of analytical reasoning~\cite{thomas2005illuminating}, where users progressively develop insights through visual exploration of data patterns, relationships, and anomalies.
This process requires sophisticated visual representations and interaction technologies that support fluid exploration while maintaining analytical rigor.

\begin{figure}[htb]
    \centering
    \includegraphics[width=\linewidth]{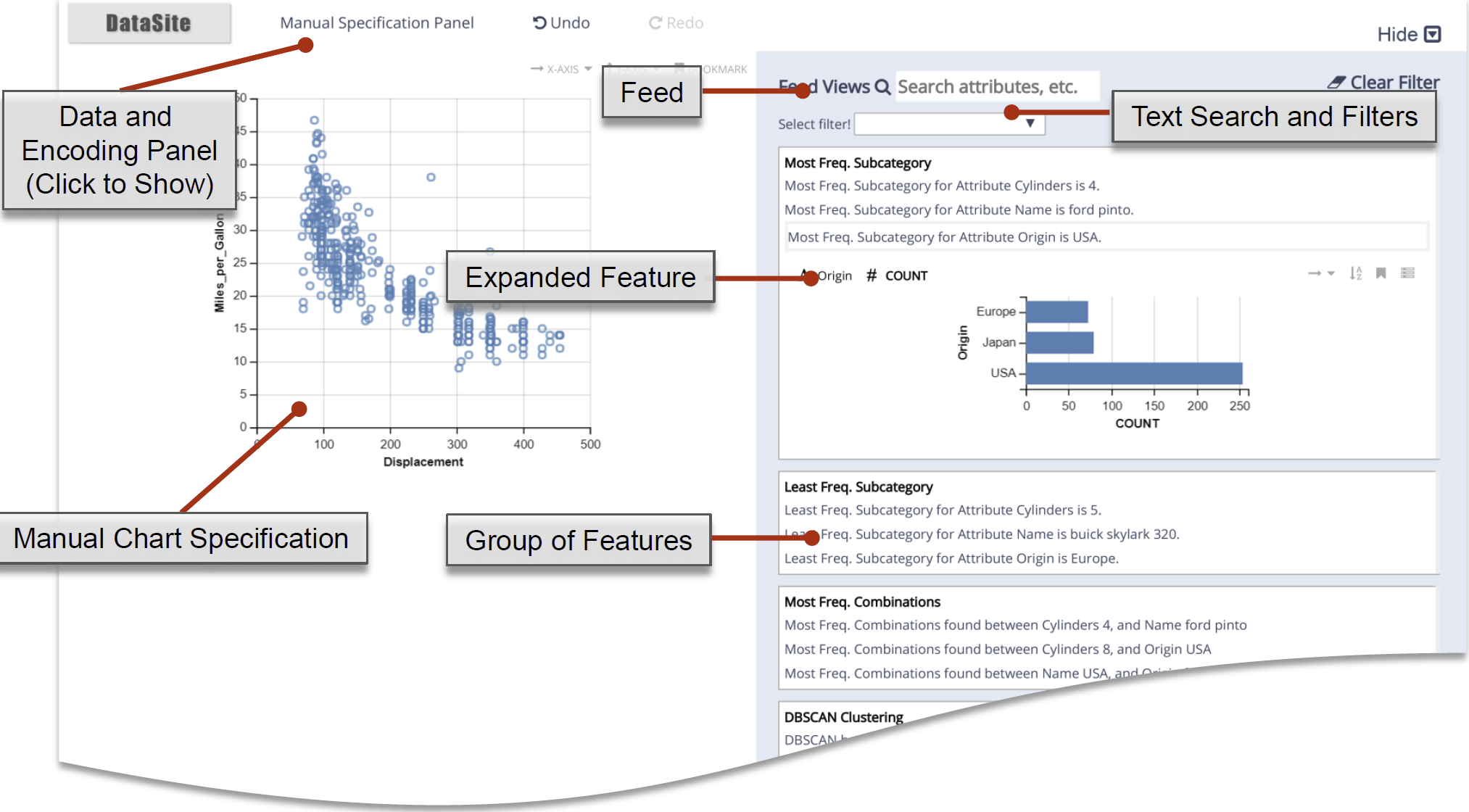}
    \caption{\textbf{DataSite.}
        This tool enhances exploratory data analysis through a human-AI collaborative framework.
        As analysts interactively develop and refine hypotheses through visual exploration, the system continuously surfaces analytically significant features in the Feed View (right), similar to a social media stream.
        This architecture maintains the analyst's agency in visual discovery while augmenting their analytical reasoning capabilities with AI-driven computational support, enabling more fluid and comprehensive exploration of data patterns, relationships, and anomalies.
    }
    \label{fig:datasite}
\end{figure}
 
\smallskip\textbf{Current HCAI Tools.}
The DataSite~\cite{DBLP:journals/ivs/CuiBYE19} platform (Figure~\ref{fig:datasite}) demonstrates how proactive computational support can enhance exploratory analysis by automatically identifying and surfacing potential patterns and relationships.
The system continuously runs multiple analysis algorithms (clustering, regression, correlation) in the background while analysts interact with visualizations, presenting findings as notifications in a social media-style feed.
Such mixed-initiative interaction~\cite{DBLP:conf/chi/Horvitz99} shares the cognitive burden between human and machine---the computational engine systematically explores combinations of variables and analysis methods, while the analyst maintains agency to investigate, verify, or dismiss suggested patterns.
Such integration of \rev{automated pattern detection with visualization} exemplifies how AI can augment human exploration capabilities without reducing analyst control.

\smallskip\textbf{Challenges, Pitfalls \& Ethical Considerations.}
In the data exploration phase, users rely on interactive visualizations to uncover patterns, which can inadvertently amplify biases, mislead interpretations, or obscure important but less prominent insights:

\begin{itemize}

\item \textit{Conformity:} Guided visual exploration may lead to all analysts following the same, standardized exploratory approach rather than letting their own expertise and knowledge guide their analysis.
Similarly, visualization recommendations tend to be deterministic, which means that they will suggest the same charts given similar data.

\item \textit{Loss of agency:} A core characteristic of visual exploration is that it is \textit{user-driven}; introducing automated guidance will necessarily diminish the analyst's agency in the exploration.

\item \textit{Cognitive atrophy:} Automated visual exploration support may yield diminished exploratory data analysis skill in the analyst.

\item \textit{Confirmation bias:} AI-guided exploration may inadvertently reinforce analysts' preexisting hypotheses by prioritizing familiar patterns over anomalous findings.

\item \textit{Pattern hallucination:} AI systems might suggest patterns in noise, leading analysts to pursue spurious correlations or non-existent trends.

\item \textit{Scalability trade-offs:} Interactive exploration of large datasets requires careful balance between responsiveness and comprehensiveness, as AI-assisted methods may sacrifice detail for speed.

\end{itemize}

\section[Schematize]{\faSitemap~SCHEMATIZE}

The schematization phase focuses on organizing discovered insights into coherent analytical structures that support understanding and reasoning.
During this phase, analysts develop mental models, establish relationships between findings, and create frameworks for understanding complex phenomena~\cite{DBLP:conf/chi/RussellSPC93}.
This process builds on foundational theories of analytical reasoning while requiring robust data representations and transformations to support progressive organization of knowledge.

\begin{figure}[htb]
    \centering
    \includegraphics[width=\linewidth]{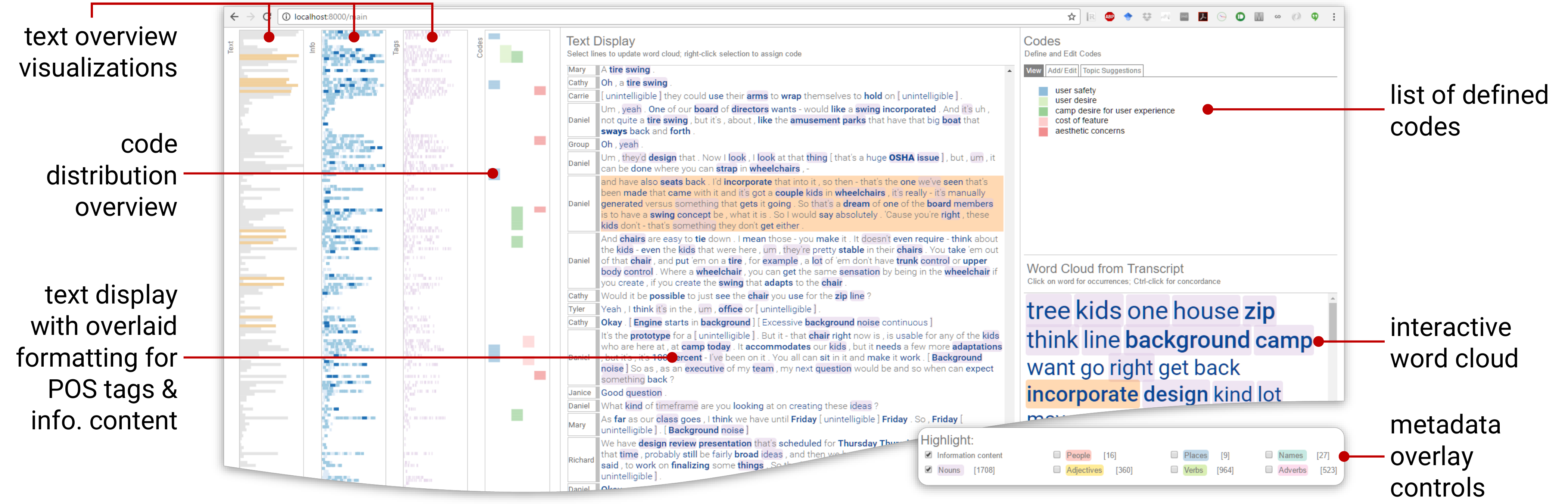}
    \caption{\textbf{Grounded theory helper.} 
        The interface supports human-AI collaborative grounded theory analysis through computational linguistics.
        Server-processed text metadata (parts of speech, entities, information content) powers coordinated visualizations that facilitate the schematization of qualitative data.
        These interactive views help analysts identify concepts and relationships while building coherent analytical structures during the iterative coding process, blending computational support with human interpretive authority.
    }
    \label{fig:gt-helper}
\end{figure}

\smallskip\textbf{Current HCAI Tools.}
Chandrasegaran et al.~\cite{DBLP:journals/cgf/ChandrasegaranB17} developed a system (Figure~\ref{fig:gt-helper}) that demonstrates how AI can support the organization of analytical insights for qualitative data analysis through multiple coordinated mechanisms.
Their system employs natural language processing for concept identification, topic modeling for thematic clustering, and information content metrics for significance detection, all integrated through interactive visualizations that help researchers progressively build coherent analytical frameworks.
The coordinated views---including structural overviews, thematic visualizations, and concept relationship displays---put analysts in the driver's seat to discover and organize patterns while maintaining agency over the schematization process.
This exemplifies how AI can augment human sensemaking by suggesting potential organizational structures while allowing flexible reorganization through interactive tools.

\smallskip\textbf{Challenges, Pitfalls \& Ethical Considerations.}
The schematization phase yields challenges in preserving context, mitigating cognitive biases, and ensuring traceability of reasoning:

\begin{itemize}

\item \textit{Accountability:} If your knowledge schemas have been constructed by an automated AI system, who is accountable for any mistakes or errors: the AI or the analyst?

\item \textit{Transparency:} Similarly, given the complexity of automated reasoning for large and/or complex datasets, how can any automated recommendations be explained to the analyst?

\item \textit{Automation bias:} Rather than formulating their own hypothesis and schemas to account for collected evidence, the analyst may choose to increasingly rely on the AI's suggestions.

\item \textit{Schema brittleness:} Automatically generated knowledge structures may fail to adapt to new evidence or edge cases; for example, a schema for financial data might break when encountering cryptocurrency transactions.

\item \textit{Context preservation:} Automated schematization may strip away important contextual information that informed the original insights; for example, temporal or situational factors that influenced data patterns.

\item \textit{Integration complexity:} Merging schemas from multiple analysts or AI systems may create inconsistencies or conflicts in knowledge representation.

\end{itemize}

\section[Report]{\faFile*~REPORT}

The reporting phase encompasses the broader goals of reaching conclusions, supporting decision-making, and effectively communicating findings.
This includes not only the production of visual artifacts and presentations but also the dissemination of results to various stakeholders with different needs and levels of expertise~\cite{thomas2005illuminating}.
The challenges in this phase span from selecting appropriate visual representations for different audiences to ensuring that conclusions are well-supported and actionable.
Success \rev{requires adopting effective presentation technologies and sustainable approaches to knowledge sharing}.

\begin{figure}[htb]
    \centering
    \includegraphics[width=\linewidth]{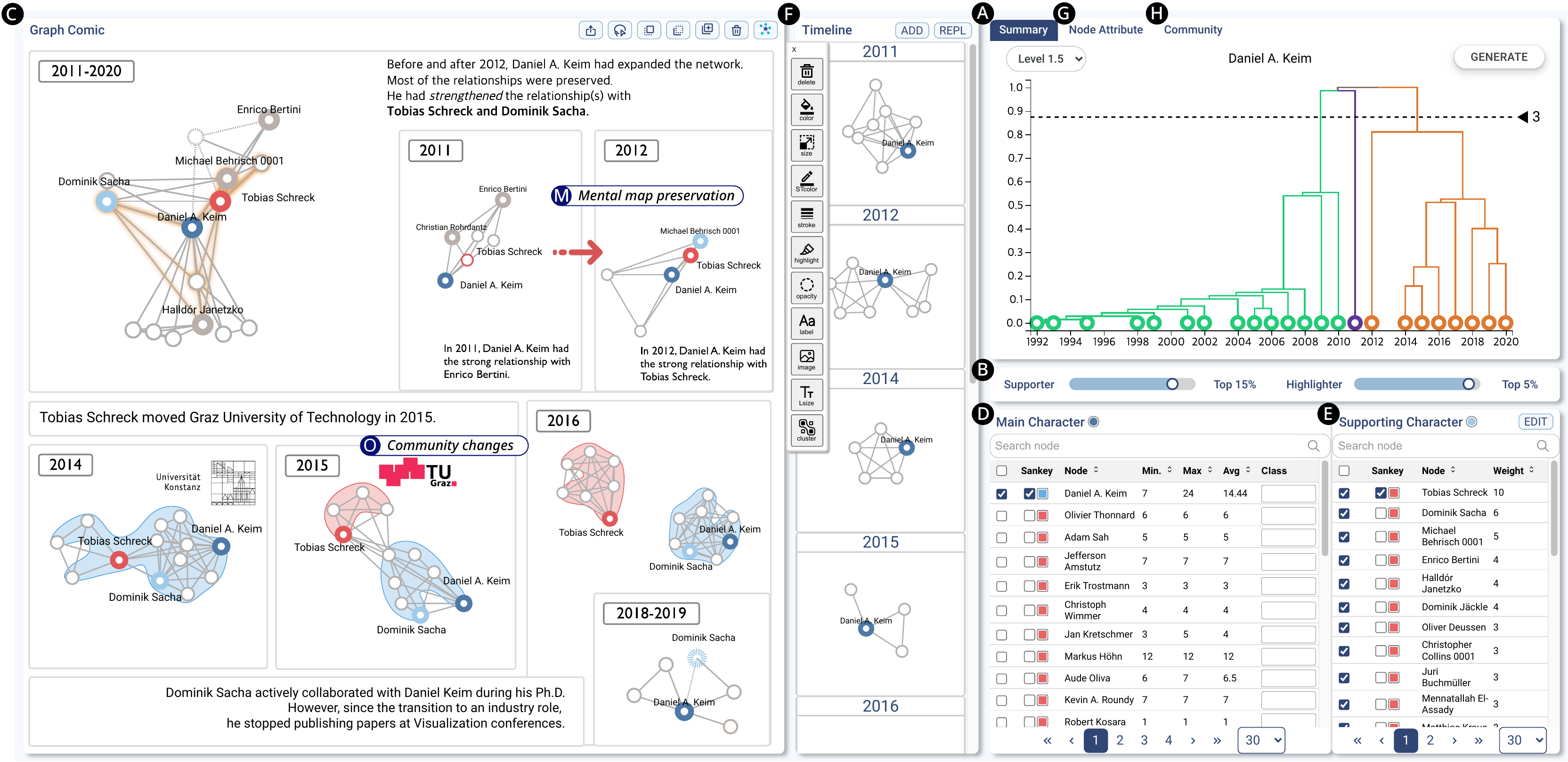}
    \caption{\textbf{DG Comics.} 
        The tool employs a human-AI collaborative approach to dynamic graph communication, featuring a summary view, interactive filtering sliders, a graph comic view, character tables, and a timeline.
        Users can access additional perspectives through a node attribute table or a community view.
        The system supports effective knowledge dissemination through mental map preservation techniques and visual community change representations, enabling stakeholders to explore actionable insights and make informed decisions.
    }
    \label{fig:dg-comics}
\end{figure}

\smallskip\textbf{Current HCAI Tools.}
The DG Comics platform~\cite{Kim2025} (Figure~\ref{fig:dg-comics}) exemplifies how semi-automated tools can support the reporting phase of visual sensemaking for complex temporal data.
The system combines AI with human creativity to help authors craft data-driven comic narratives from dynamic graphs.
It automatically segments temporal events using hierarchical clustering while preserving causality, identifies key actors and relationships through graph metrics, and provides an integrated authoring environment where users can construct their stories.
This balanced approach to automation---where algorithms handle the complex task of temporal segmentation while preserving human agency in narrative choices---demonstrates how visualization-enabled HCAI tools can augment traditional visualization workflows for storytelling.

\smallskip\textbf{Challenges, Pitfalls \& Ethical Considerations.}
The reporting and presentation phase demands careful consideration of automation, narrative coherence, and the ethical implications of framing and emphasizing specific results:

\begin{itemize}

\item \textit{Automation bias:} Automated support for presentation may yield undue reliance on the AI's suggestions over one's own conclusions and preferences; \rev{e.g.}, accepting \rev{AI interpretations} of statistical findings without critical evaluation.

\item \textit{Conformity:} Blindly accepting automated presentation suggestions may lead to \rev{homogenized data-driven stories}, potentially obscuring unique insights or domain-specific nuances.

\item \textit{Loss of agency:} Automation may diminish analysts' willingness to shape narratives, particularly when AI-generated presentations appear polished, authoritative, and backed by strong (if perhaps circumstantial) evidence.

\item \textit{Fairness and framing:} All narratives inherently contain perspective bias; \rev{automated reports may inadvertently} amplify certain viewpoints or underrepresent minority perspectives in the data.

\item \textit{Context translation:} Automated reporting may fail to adequately translate technical findings for different audience levels; for example, oversimplifying complex statistical relationships for executive summaries.

\item \textit{Reproducibility:} AI-generated reports may obscure the analytical process, making it difficult for others to \rev{verify the findings independently.}

\item \textit{Stakeholder engagement:} Over-reliance on automated reporting may reduce opportunities for \rev{stakeholder input and refinement of insights.}

\end{itemize}

\section[Updating the R\&D Agenda]{UPDATING THE R\&D AGENDA}

The fundamental strength of visualization lies not in answering questions, but in helping analysts formulate the questions they didn't know to ask; what Thomas and Cook call ``\textit{discover the unexpected.}''
Such exploratory data analysis (EDA) is even more important today, where complex AI models and massive datasets require novel approaches to understanding and validation, than it was 20 years ago.
While AI systems excel at providing answers to well-formed queries, they often struggle with the open-ended exploration and hypothesis generation that visualization enables.

As we look ahead to the next two decades of visual analytics research, we envision a convergence between visual analytics and artificial intelligence.
This convergence isn't about AI replacing human visual analysis, but rather about creating systems that combine the computational power of AI with the irreplaceable human capacity for exploratory analysis---finding patterns, generating hypotheses, and closing the analytical loop. 
This fusion will democratize data analysis capabilities while maintaining human agency and accountability in the analytical process.

\subsection{Advancing AI through Visual Analytics}

\rev{We should not forget that visual representations can be a powerful enabler for AI as well.
Therefore, the visualization community must take a proactive role in making AI systems more transparent and interactive, essentially turning pipelines into feedback loops.
For example, visual analytics can significantly advance explainable AI (XAI)~\cite{DBLP:journals/air/MinhWLN22} by providing interactive tools for understanding model behavior, tracking decision processes, and validating outcomes.}

As AI systems become more complex and pervasive, visualization will also play an increasingly important role in maintaining human oversight.
We need visual interfaces that work across different model architectures and support both cloud-based and local processing paradigms, ensuring flexibility and data privacy.
These tools must evolve beyond simple model interpretation to support comprehensive understanding of AI systems' behavior in complex, real-world contexts.

\subsection{\rev{Democratizing Visual Analytics \& AI}}

The convergence of AI and visualization creates unprecedented opportunities to democratize data analysis for everyone.
Just like how tools such as Copilot have revolutionized software engineering, AI can reduce barriers to visualization authoring and customization while maintaining human agency in the analytical process.
This includes quality assurance for detecting misleading visualizations, tools for adapting visualizations across different visualization literacy levels, and support for local, privacy-preserving data analysis.
\rev{This will, in effect, diminish the distinction between consumers, creators, and developers of visualizations.
Practically speaking, it will allow AI to enhance the development of visualization literacy by aiding both consumption---such as detecting misleading graphs---and production, by actively suggesting techniques, improvements, or alternatives.}

\rev{Democratization also means supporting people being informed, taking ownership, and learning the skills necessary for their work.
This could mean using a progressive form of automation where the AI serves as ``training wheels'' for data analysis, gradually teaching the users these skills and improving their understanding as they proceed with the goal of eventually reducing the automation to almost nothing.
In the end, it is the difference between feeding someone for a day and teaching them to fish for themselves.}

\rev{The ultimate goal} isn't necessarily to turn everyone into a data scientist, but rather to make sophisticated data analysis accessible to domain experts across fields.
\rev{This means developing systems that adapt to users' visualization literacy levels, provide appropriate guidance without being intrusive, and support customization for specific domains.}
\rev{The goal is to maintain high accuracy and interpretability even for non-experts.}

\subsection{\rev{Future Innovation}}

To realize this vision, we need advances in both technical capabilities and scientific methodologies.
On the technical side, we need robust, secure frameworks for visual model understanding, scalable visual exploration techniques that support both local and cloud-based processing while preserving privacy, and comprehensive systems for model accountability.

The methodological challenges are equally significant.
We need new approaches to evaluate long-term impact on analytical processes, security implications of AI-enhanced visualization systems, and the effectiveness of privacy-preserving visual analytics.
These methodologies must consider both immediate task completion and longer-term effects on analytical capabilities and domain understanding.
Furthermore, we need frameworks for measuring how visualization tools support accountability in AI development and deployment across different organizational contexts.

\rev{This paper has primarily focused on \textit{what} AI can contribute to visual analytics, and less \textit{how} this should be realized in terms of UI and interaction design.
The chat has become the de facto interface for AIs, but this interaction has clear shortcomings for visual communication.
Exploring multimodal AI interaction for visual analytics remains a vast research frontier, intersecting all four capabilities and sensemaking phases.
Here, the past decades of HCI research provides the necessary foundation for designing effective and usable HCAI, leveraging established principles in usability, managing the cognitive load, supporting learnability, and more.}

\subsection{Preparing for the Next Two Decades}

The path ahead requires careful attention to both technical and social considerations.
The visualization community must develop open, modular tools that work across different AI ecosystems while maintaining security and privacy.
We must balance the democratization of visualization capabilities with the need for quality assurance and appropriate human oversight.

\rev{Ethical, safe, and responsible AI is a particularly important future consideration where visualization can play a fundamental role.
We believe that interactive visual representations present the best path to an explainable approach to AI that can help keep human users informed, engaged, and in control of AI systems.}

\rev{A truly illuminated future will mean} creating systems that reduce barriers to visualization while maintaining rigorous standards for accuracy and interpretability.
It means developing tools that support both novice and expert users without compromising analytical power.
Most importantly, it means ensuring that as AI systems become more sophisticated, humans retain meaningful agency~\cite{Shneiderman2022} in the analytical process through powerful, accessible visual interfaces.

\def\refname{REFERENCES}


\begin{IEEEbiography}{Niklas Elmqvist} is a Villum Investigator and professor in the Department of Computer Science at Aarhus University in Aarhus, Denmark.
His research interests include data visualization, human-computer interaction, and human-centered AI.
He received the Ph.D.\ degree in 2006 in computer science from Chalmers University of Technology.
He is a Fellow of the IEEE and the ACM.
Contact him at elm@cs.au.dk.
\end{IEEEbiography}

\begin{IEEEbiography}{Clemens Nylandsted Klokmose} is an associate professor in the Department of Computer Science at Aarhus University in Aarhus, Denmark.
His research interests include collaboration, user interface technologies, and human-computer interaction.
He received the Ph.D.\ degree in 2009 in computer science from Aarhus University.
Contact him at clemens@cs.au.dk.
\end{IEEEbiography}

\end{document}